\documentclass[twocolumn,showpacs,preprintnumbers,amsmath,amssymb,superscriptaddress,groupedaddress,prmater]{revtex4}

\usepackage{graphicx}
\usepackage{amsmath}
\usepackage{amssymb}
\usepackage{amsfonts}
\usepackage{dcolumn}
\usepackage{epsfig}
\usepackage{subfigure}
\usepackage{bm}

\begin{document}

\title{\bf Nonmonotonic magnetic field dependence of remanent ferroelectric polarization in reduced-graphene-oxide-BiFeO$_3$ nanocomposite}

\author{Tania Chatterjee}
\affiliation{Advanced Materials and Chemical Characterization Division, CSIR-Central Glass and Ceramic Research Institute, Kolkata 700032, India}
\author{Arnab Mukherjee}
\affiliation{Functional Materials and Device Division, CSIR-Central Glass and Ceramic Research Institute, Kolkata 700032, India}
\author{Prabir Pal} \affiliation{Energy Materials and Device Division, CSIR-Central Glass and Ceramic Research Institute, Kolkata 700032, India}
\author{S.D. Kaushik} \affiliation{UGC-DAE Consortium for Scientific Research, Bhabha Atomic Research Centre, Mumbai 400085, India}
\author{V. Siruguri} \affiliation{UGC-DAE Consortium for Scientific Research, Bhabha Atomic Research Centre, Mumbai 400085, India} 
\author{Swarupananda Bhattacharjee}
\affiliation{School of Materials Science and Nanotechnology, Jadavpur University, Kolkata 700032, India}
\author{Chandan Kumar Ghosh} 
\affiliation{School of Materials Science and Nanotechnology, Jadavpur University, Kolkata 700032, India}
\author{Dipten Bhattacharya} \email{dipten@cgcri.res.in} 
\affiliation{Advanced Materials and Chemical Characterization Division, CSIR-Central Glass and Ceramic Research Institute, Kolkata 700032, India}

\date{\today}

\begin{abstract}
In a nanocomposite of reduced graphene oxide (RGO) and BiFeO$_3$ (BFO), the remanent ferroelectric polarization is found to follow nonmonotonic magnetic field dependence at room temperature as the applied magnetic field is swept across 0-20 kOe on a pristine sample. The remanent ferroelectric polarization is determined both from direct electrical measurements on an assembly of nanoparticles and powder neutron diffraction patterns recorded under 0-20 kOe field. The nanosized ($\sim$20 nm) particles of BFO are anchored onto the graphene sheets of RGO via Fe-C bonds with concomitant rise in covalency in the Fe-O bonds. The field-dependent competition between the positive and negative magnetoelectric coupling arising from magnetostriction due to, respectively, interface and bulk magnetization appears to be giving rise to the observed nonmonotonic field dependence of polarization. The emergence of Fe-C bonds and consequent change in the magnetic and electronic structure of the interface region has influenced the coupling between ferroelectric and magnetic properties remarkably and thus creates a new way of tuning the magnetoelectric properties via reconstruction of interfaces in nanocomposites or heterostructures of graphene/single-phase-multiferroic systems.  
\end{abstract}

\pacs{75.80.+q, 75.75.+a, 77.80.-e}
\maketitle

\section{Introduction}

The reconstruction of crystallographic, magnetic, and electronic structures at the interface turns out to be quite an effective way of preserving the multiferroic orders and augmenting the coupling among the order parameters in multiferroics based heterostructures/composites \cite{Huang}. Among different heterostructures or composites, graphene/BiFeO$_3$ or reduced-graphene-oxide-BiFeO$_3$ systems have attracted a lot of attention \cite{Zanolli-1,Song,Zanolli-2,Wu,Barman}. Large exchange field $B_{ex}$ (of the order of ten to hundred tesla) in graphene/BiFeO$_3$ heterostructure induces proximity effect driven magnetism \cite{Song,Wu} while, in graphene/BaMnO$_3$ system, the magnetic structure of BaMnO$_3$ exhibits change from antiferromagnetic to ferromagnetic orientation at the interface \cite{Zanolli-2}. It has also been shown \cite{Volonakis} that graphene-methylammonium lead iodide (CH$_3$NH$_3$PbI$_3$) composite exhibits ferroelectricity because of reconstruction of crystallographic structure in the interface regions. Given all these results, it is important to examine how reconstruction of magnetic and electronic structures by exchange coupled surface atoms of graphene and nanoscale BiFeO$_3$ influences the multiferroicity in BiFeO$_3$. In this paper, we show that the reduced-graphene-oxide-BiFeO$_3$ (RGO/BFO) nanocomposite exhibits significant decrease in ferromagnetic component yet enhancement in coercivity and remarkable nonmonotonic variation of ferroelectric polarization with applied magnetic field ($H$). Bulk BiFeO$_3$ exhibits \cite{Lee} suppression of off-centering of Fe ions [and consequently, suppression of ferroelectric polarization ($P$)] below $T_N$ (off-centering of Bi ions remains unaffected) since it is driven by negative magnetostriction within Fe-O-Fe spin structure. It also yields monotonic suppression of $P$ under $H$. The $P$ arising out of magnetostriction due to the magnetization in the interface region of the RGO/BFO nanocomposite, on the other hand, could exhibit monotonic rise with $H$. Field-dependent competition between these two opposite trends (emerging within bulk and interface spin structures) possibly yields this nonmonotonic $H$ dependence of $P$ in the nanocomposite.      

\section{Experimental Details}

The RGO/BFO nanocomposite has been prepared by hydrothermal technique. Graphene oxide (GO) was prepared from graphite via modified Hummers method \cite{Hummers} in which 70 ml of sulphuric acid (H$_2$SO$_4$) was taken within an ice bath (for maintaining the temperature below 20$^o$C) and 1g of graphite powder and 0.5g of sodium nitrate (NaNO$_3$) were added to the acid solvent and stirred for a while. After that, 3g of potassium permangante (KMnO$_4$) was added gradually and stirred vigorously. We allowed the bath temperature to stabilize and then removed the ice bath to attain room temperature. The mixed solution was stabilized at room temperature for 2h and then diluted by adding 5\% hydrogen peroxide (H$_2$O$_2$) solution slowly till the color turned brownish yellow. It confirmed complete oxidation of graphite. Finally, the solution was stirred for 30 minutes to further exfoliate the as-formed graphite oxide (GO). The GO, thus synthesized, was collected by filtration and washing for several times with 1N hydrochloric acid (HCl) and deionized water. Synthesized GO and BFO precursors - [Bi(NO$_3$)$_3$,5H$_2$O] and Fe(NO$_3$)$_3$,6H$_2$O] - were used for the hydrothermal synthesis \cite{Li} of the composite. The aqueous solution of GO was reduced to form reduced graphene oxide (RGO) using ascorbic acid as the reducing agent. The 8M potassium hydroxide (KOH) and the precursors of BFO were added in stoichiometric ratio. Once homogenized, the as-prepared RGO solution was added to the mixture to maintain 1.50 weight\% of RGO in the resulting mixture. The ammonium hydroxide (NH$_4$OH) was added in 1.0 weight\% to facilitate the formation of BiFeO$_3$ nanoparticles embedded in the RGO matrix. The solution was stirred for 30 minutes and put into an autoclave for hydrothermal treatment at 170$^o$C for 2, 4, 5, 6 hours. The sample treated hydrothermally for 6h turned out to be phase pure and was designated as Com-H. The nanocomposite Com-H was further heated to 550$^o$C for 20 minutes. This treatment removes the carbon skeleton and leaves the bare nanoparticles of BiFeO$_3$ (BFO-H). The thermogravimetric analysis was carried out under both air and inert (Ar) atmosphere to track the formation of BFO-H.

All the samples were characterized by powder X-ray diffraction (XRD), infrared (IR) and Raman spectrometry, scanning and tunneling electron microscopy (SEM and TEM), and x-ray photoelectron spectroscopy (XPS). The XPS measurements were performed by PHI 5000 VERSAPROBE II, Physical Electronics System, equipped with monochromatic Al $k_{\alpha}$ (1486.7 eV) focussed x-ray source and a multi-channeltron hemispherical electron energy analyzer. All the spectra were collected at an emission angle of 45$^o$ with the base vacuum of 5.0 $\times$ 10$^{-10}$ mbar. The binding energies were referenced by measuring C 1s and keeping it at 284.6 eV. The total energy resolution was estimated to be $\sim$400 meV for monochromatic Al $k_{\alpha}$ line with pass energy 11.750 eV. A charge neutralizer was used to compensate the surface charging of the samples. A background has been subtracted from the measured raw data. The magnetic hysteresis loops were measured by LakeShore Vibrating Sample Magnetometer (VSM; Model 7407) under $\pm$20 kOe field at room temperature while the ferroelectric polarization was measured by the Precision LC-II (Radiant Technologies Inc.) ferroelectric loop tester. For preparing the samples for electrical measurements, nanoparticles were dispersed within ethanol and spin coated onto the Si/SiO$_2$ substrate to form the films \cite{supplementary} (thickness of the film was $\sim$10 nm as measured by ellipsometry). Silver dots were used as electrodes. The powder neutron diffraction patterns were recorded at room temperature at the PD-3 beamline of National Facility for Neutron Beam Research (NFNBR), Dhruva Reactor, Mumbai using a monochromatic beam of wavelength 2.315 \AA. The magnetic field was varied across 0-20 kOe.

\begin{figure}[h!]
 \begin{center}
    \includegraphics[scale=0.20]{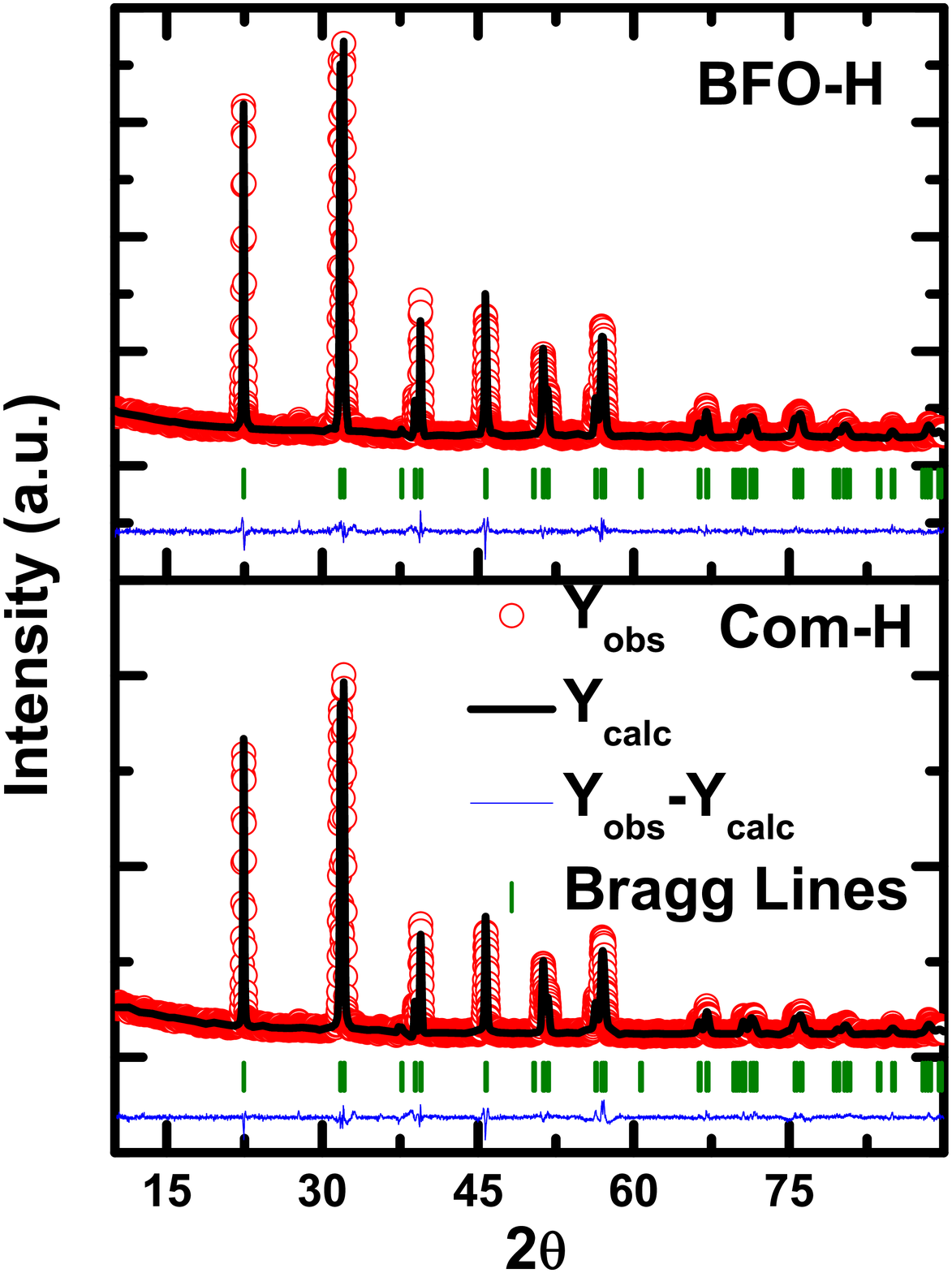}
    \end{center}
\caption{The room temperature x-ray diffraction data and their refinement.}
\end{figure}

\begin{figure}[h!]
 \begin{center}
  \subfigure[]{\includegraphics[scale=0.15]{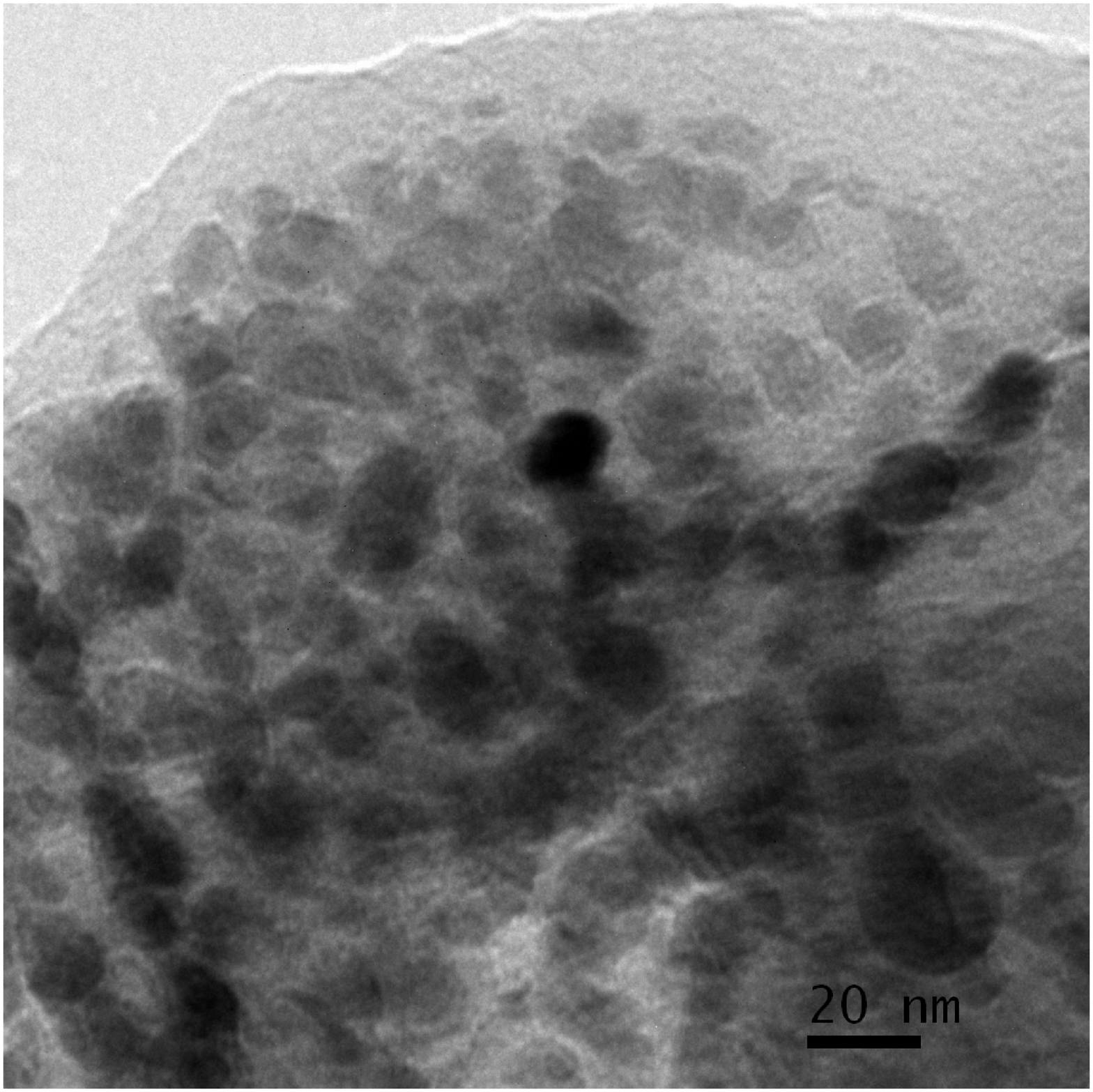}} 
   \subfigure[]{\includegraphics[scale=0.15]{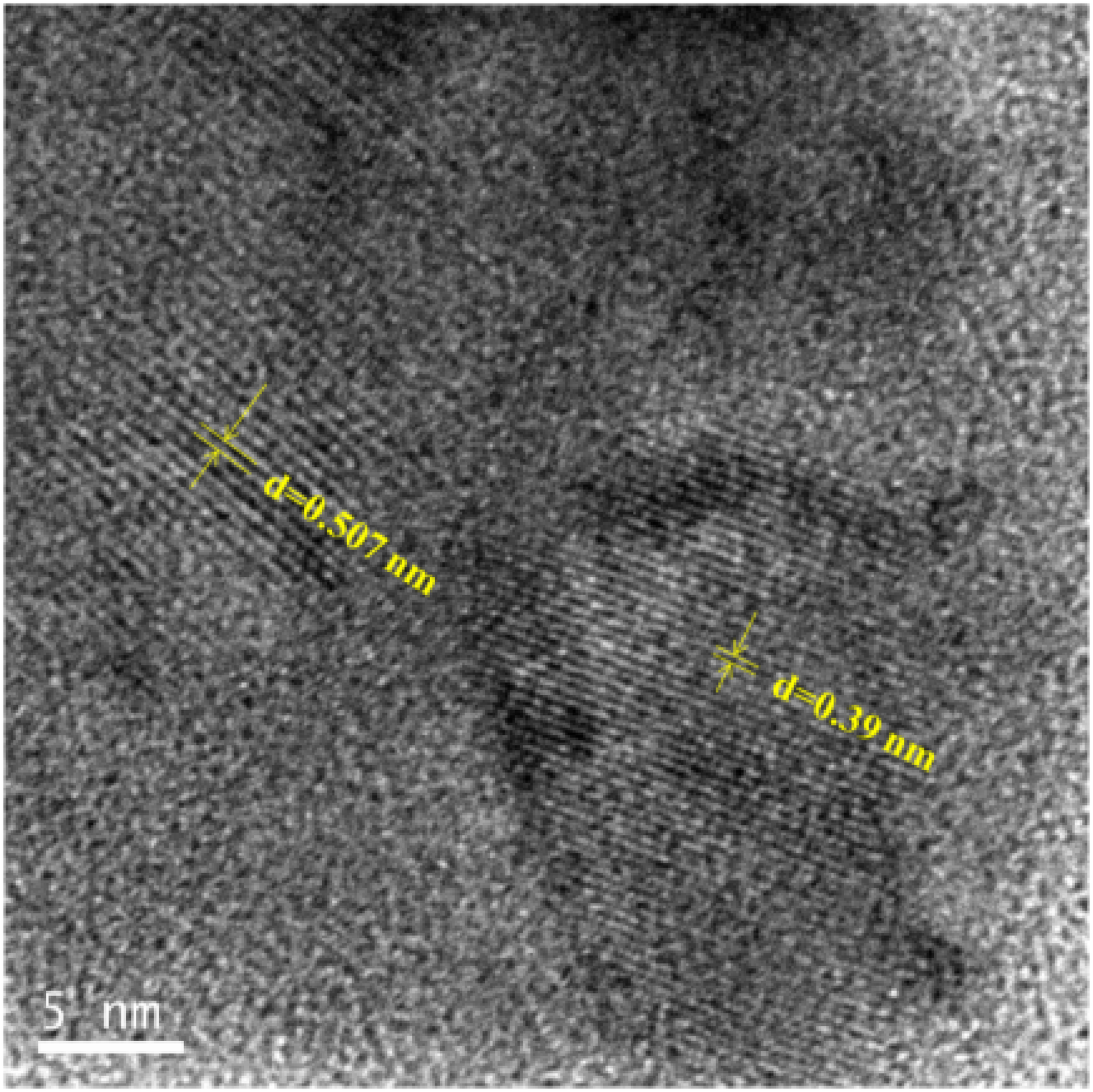}}
    \end{center}
\caption{The (a) bright field TEM and (b) high resolution TEM (HRTEM) images of the BFO/RGO nanocomposite.}
\end{figure}

\begin{figure}[h!]
\begin{center}
   \subfigure[]{\includegraphics[scale=0.30]{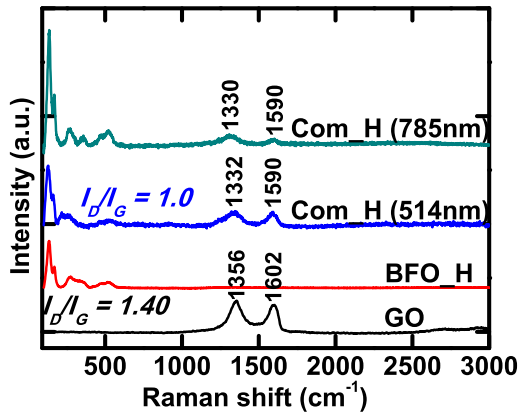}}
   \subfigure[]{\includegraphics[scale=0.30]{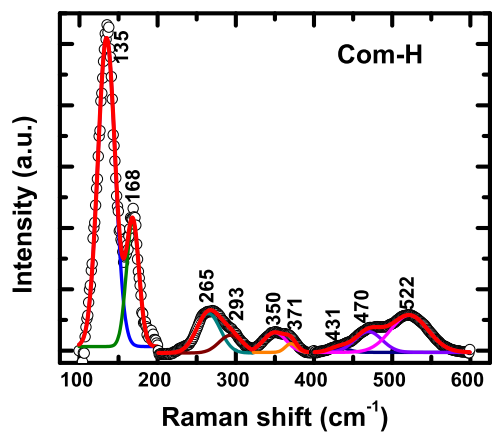}}
   \subfigure[]{\includegraphics[scale=0.30]{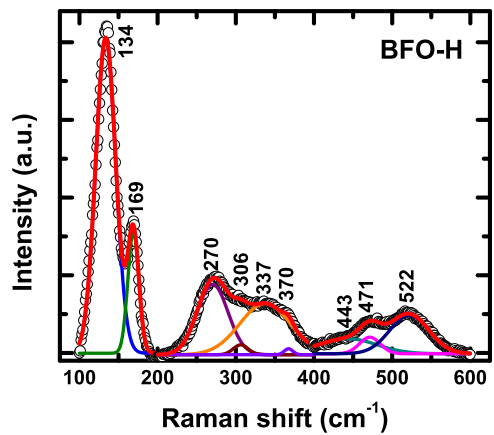}}
   \end{center}
\caption{Raman spectra for the composite (COM-H), pure BFO (BFO-H), and GO systems.}
\end{figure}

\section{Results and Discussion}

Figure 1 shows the XRD data for Com-H and BFO-H. The data were refined by FullProf. Both Com-H and BFO-H were found to assume $R3c$ space group. The average global crystallographic structure has not changed in the nanocomposite. BiFeO$_3$ assumes $R3c$ symmetry due to the presence of both polar distortion as well as antiferrodistortive rotation of FeO$_6$ octahedra (which was shown \cite{Spaldin} to be related to the magnetization) around the axis of polarization [111]. The detailed results of the refinement (lattice parameters, ion positions, bond lengths and angles) including the fit statistics are given in the supplementary document \cite{supplementary}. Figures 2(a),(b) show the bright field TEM and high resolution TEM (HRTEM) images for Com-H. The image analysis yields the average size of BFO particles to be $\sim$18-20 nm which is comparable to the particle size of BFO-H (shown in \cite{supplementary}). A certain fraction of the BFO particles are anchored onto the RGO layers. The HRTEM image shows the interface between (012) plane of BFO (d = 0.39 nm) particle and the RGO layer with interlayer spacing 0.49 nm. It indicates deposition of graphene layer onto the (100)$_c$ plane of BiFeO$_3$. The interlayer spacing of RGO is much larger than that in graphite (d = 0.37 nm) due to exfoliation. 

The Raman spectrometry of the Com-H and BFO-H are shown in Figs. 3(a),(b),(c). The experiments were carried out with both 514 nm and 785 nm He-Ne laser excitations. The 514 nm laser probes the organic compound while the inorganic system is probed by 785 nm laser. It is found that the intensity ratio of the characteristic D and G peaks at 1355 cm$^{-1}$ and 1593 cm$^{-1}$ - which, respectively, correspond to the dangling bonds from defects containing sp$^3$ hybridized orbitals and planar vibration of C ions bonded via sp$^2$ hybridized orbitals - is 1.00 in the Com-H. Using the relations \cite{Jorio,Eng} $L_D^2 = [(1.8 \pm 0.5)\times 10^{-9} \times \lambda^4]/[I_D/I_G]$ and $n_D = 10^{14}/\pi L_D^2$, where $L_D$ is the effective distance between the defects (in nm), $n_D$ is the defect concentration (in cm$^{-2}$), $\lambda$ is the wavelength of the excitation, and $I$ is the intensity of the D or G peaks, we find that the defect concentration ($n_D$) and the spacing between the defects ($L_D$) in the RGO layers are 1.98 $\times$ 10$^{27}$ cm$^{-2}$ and 12.67 nm, respectively. The corresponding figures for the graphite oxide (GO) are 2.77 $\times$ 10$^{27}$ cm$^{-2}$ and 10.70 nm, respectively. The decrease in defect concentration and increase in defect spacing in Com-H with respect to GO indicates lesser extent of functionalization of the graphene layer in Com-H. This observation could have some significance in pointing out that the BiFeO$_3$ nanoparticles used for functionalization of the graphene results in coexistence of both bonded and nonbonded nanoparticles. 

The characteristic modes of BFO exhibits substantial red and blue shifts in the Com-H. According to group theory, there are 13 Raman active modes of BFO with the rhombohedral $R3c$ structure which are defined by the following irreducible representations \cite{Hermet} $\Gamma_{Raman, R3c}$ = 4$A_1$ $\bigoplus$ 9$E$; where 4 $A_1$ modes are polarized along z axis and the rest 9 $E$ doubly degenerate modes are polarized along x-y plane. In BFO-H and Com-H (Figs. 3b and 3c), nine active modes could be observed out of the 13 modes. The lineshape of the peaks was fitted by Gaussian function. Their assignment \cite{Hermet,Scott} together with the modes calculated from first-principles \cite{Hermet} are listed in Table I.

\begin{table}[ht]
{\caption {The Raman modes for Com-H and BFO-H samples.}

\begin{tabular}{p{1.0in}p{0.6in}p{0.6in}p{0.6in}} \hline\hline
Raman modes & BFO-H (cm$^{-1}$) & Com-H (cm$^{-1}$) & Calculated (cm$^{-1}$)  \\ \hline
$E$(TO) & 133.5 & 134.2 & 152\\
$A_1$(TO) & 169.3 & 168.7 & 167\\
$A_1$(LO) \newline $A_1$(TO) & 270.3 & 265.4 & 277 \newline 266\\
$A_1$(TO) & 306.2 & 293.1 & 318\\
$E$(LO) \newline $E$(TO) & 337.5 & 350.7 & 332 \newline 335\\
$E$(TO) \newline $E$(LO) & 369.7 & 371.1 & 378 \newline 386\\
$E$(TO) \newline $E$(LO) & 443.4 & 431.2 & 409 \newline 436\\
$A_1$(LO) & 471.2 & 470.1 & 509\\
$A_1$(TO) & 522.1 & 522.4 & 517\\
\hline \hline

\end{tabular}}
\end{table}

\begin{figure}[h!]
\begin{center}
   \subfigure[]{\includegraphics[scale=0.30]{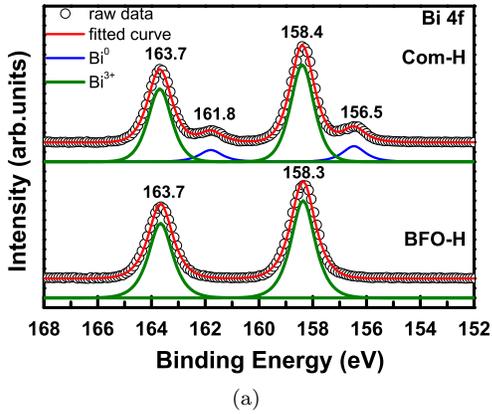}}
   \subfigure[]{\includegraphics[scale=0.30]{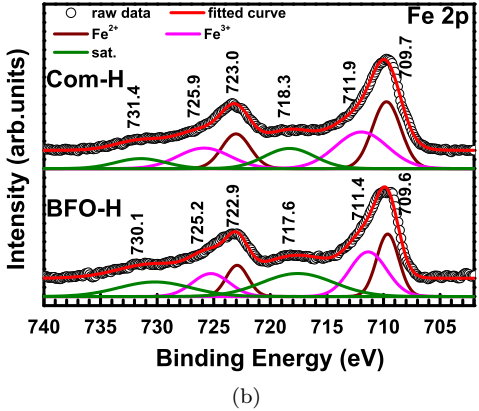}}
   \end{center}
\caption{X-ray photoelectron spectra and their fitting for (a) Bi 4f and (b) Fe 2p in pure BFO (BFO-H) and composite (COM-H) systems.}
\end{figure}

\begin{figure}[h!]
 \begin{center}
    \includegraphics[scale=0.30]{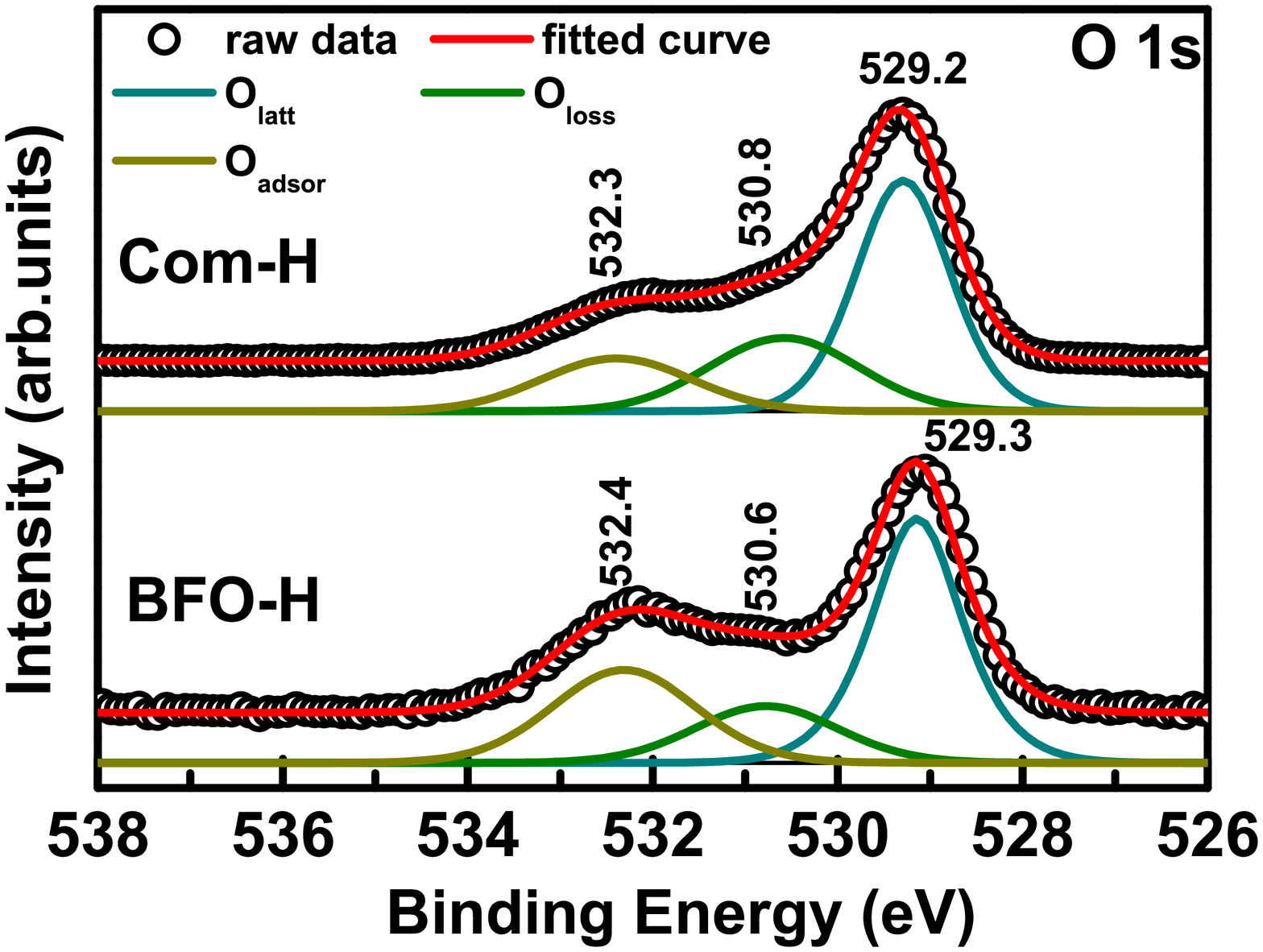}
    \end{center}
\caption{X-ray photoelectron spectra and their fitting for C 1s in pure BFO (BFO-H) and composite (COM-H) systems.}
\end{figure}

\begin{figure}[h!]
 \begin{center}
    \includegraphics[scale=0.30]{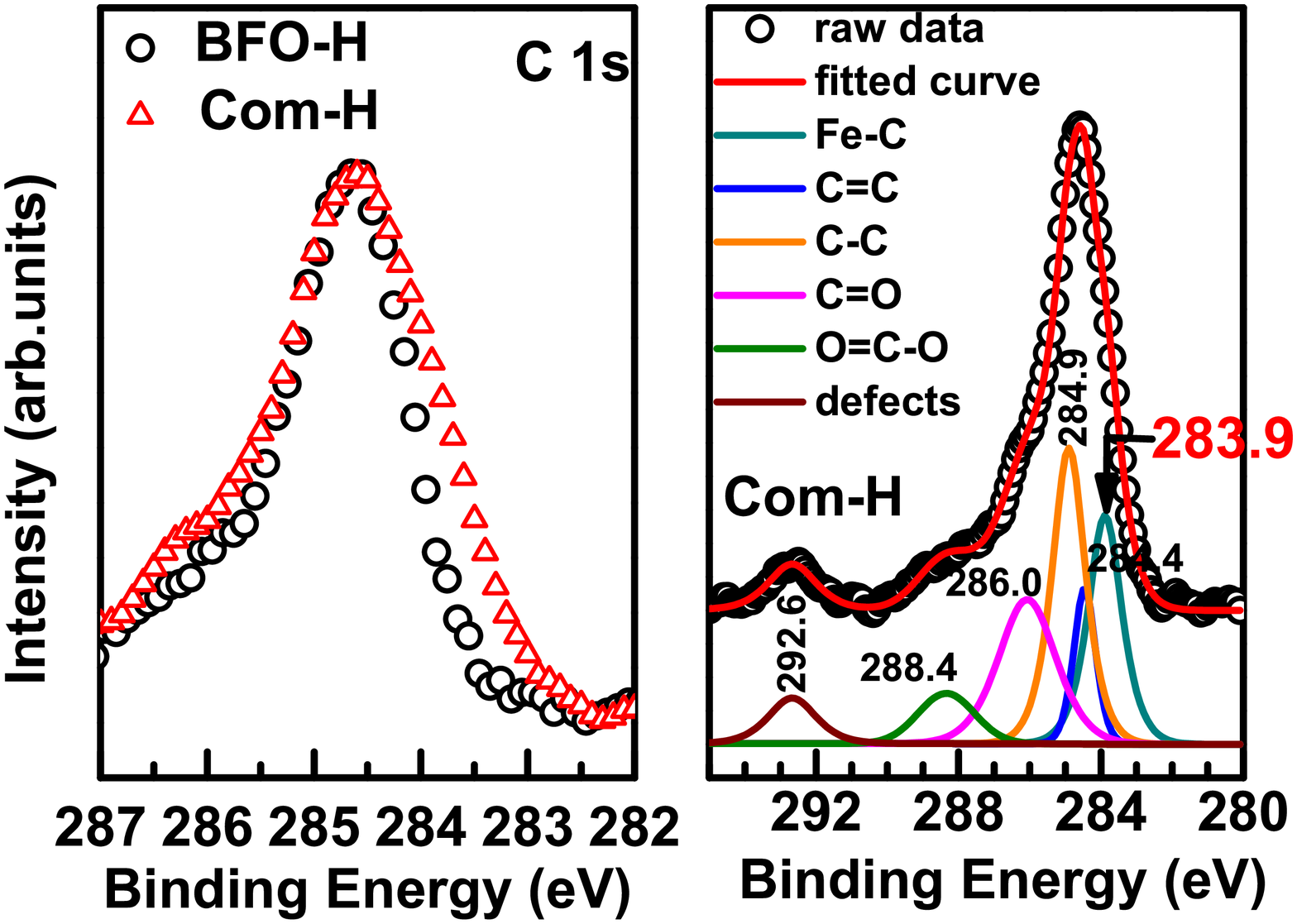}
    \end{center}
\caption{X-ray photoelectron spectra and their fitting for O 1s in pure BFO (BFO-H) and composite (COM-H) systems.}
\end{figure}

\begin{figure}[h!]
 \begin{center}
    \includegraphics[scale=0.30]{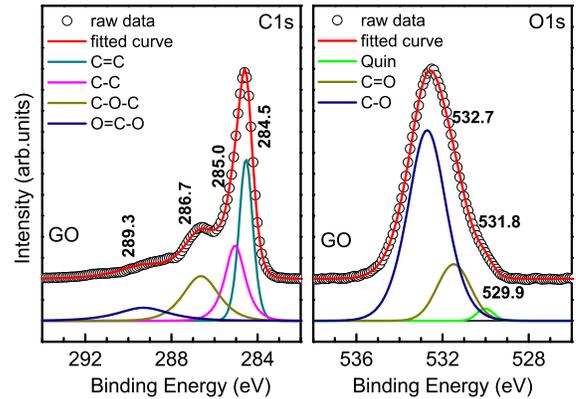}
    \end{center}
\caption{High resolution C 1s and O 1s X-ray photoelectron spectra of GO highlighting presence of different species on its surface.}
\end{figure}

The lower order $A_1$ and $E$ modes are related to the Bi-O and Fe-O vibration while the higher order peaks arise due to the change in the oxygen motion states. Comparison of the Raman spectra observed in BFO-H and Com-H (Fig. 3) reveals that (i) the lower order modes (till 167 cm$^{-1}$), influenced by Bi atoms, do not exhibit any substantial shift while (ii) the modes influenced by Fe ions and oxygen motion states are either red or blue shifted. For example, the $A_1(LO)$/$A_1(TO)$ and $E$(TO) modes at $\sim$270 cm$^{-1}$, $\sim$306 cm$^{-1}$, and $\sim$443 cm$^{-1}$ are redshifted by 5-13 cm$^{-1}$ while the $E(LO)/E(TO)$ one at $\sim$337.5 cm$^{-1}$ is blueshifted by $\sim$13 cm$^{-1}$ in Com-H. This could be due to the change in the Fe-O bond lengths and the motion states of oxygen. The formation of Fe-C bonds at the BFO/RGO interface (discussed later) could have some influence on the bulk lattice and its distortion. The distortion in the Fe-O bonds have contributions from both polar distortion and antiferrodistortive tilt. The calculation of the Fe-O bond length (supplementary data\cite{supplementary}) shows that indeed the difference in the bond lengths has decreased in Com-H. Similar observation has earlier been made by others \cite{Alonso} in different perovskite oxide systems. The infrared spectra too corroborate these observations together with the confirmation of formation of RGO and are included in the supplementary document \cite{supplementary}.

The high resolution XPS data for Bi 4f and Fe 2p are shown in Figs. 4(a) and 4(b), respectively. The Bi 4f$_{7/2}$ and 4f$_{5/2}$ spin-orbit doublet peaks are located at 158.4 and 163.7 eV, respectively, which are found to correspond to Bi$^{3+}$ states from Bi-O bonds. Interestingly, the binding energy of Bi$^{3+}$ states remains same in both the BFO-H and Com-H samples. It indicates that no substitution has taken place at the Bi$^{3+}$ sites in the Com-H sample. Of course, an additional weak pair of doublet peaks was observed in Com-H in the lower binding energy side which points to the presence of Bi$^0$ states. To estimate the percentage of Bi$^0$ states in the Com-H sample, we have adopted the $\chi^2$ iterative fit of the Bi 4f core-level region using two pair of doublet components corresponding to the Bi$^{3+}$ (grey) and Bi$^0$ (blue) states, as shown in Fig. 4(a). Except for the peak area, the energy positions, line shapes, widths, and all other parameters were kept same during the fitting of two pairs of doublet peaks. The Bi$^0$ states 4f$_{7/2}$ and 4f$_{5/2}$ (spin-orbit doublet peaks) are centered at $\sim$156.5 and $\sim$161.8 eV, respectively. The relative percentage of Bi$^0$ states is found to be $\sim$14.3\%. 

The Fe2p$_{3/2}$ and Fe2p$_{1/2}$ spin-orbit doublet peaks are located around 709.6 eV and 722.9 eV, respectively, with a pair of shake-up satellite peaks \cite{Pal} located at 8.0 eV above their spin-orbit doublet peaks. We have carefully fitted the spectra using the following fitting parameters for the Fe$^{2+}$: spin-orbit splitting 13.3 eV, branching ratio 2.0, full widths at half maximum (FWHM) 2.3 eV; an integral background was subtracted before fitting. Similarly, the spectra for Fe$^{3+}$ were fitted by considering spin-orbit splitting 13.8 eV, branching ratio 2.0, and FWHM 3.8 eV. The fitted Fe 2p spectra highlight the characteristic doublet peaks of Fe$^{2+}$-O species at $\sim$709.6 eV and $\sim$722.9 eV for Fe2p$_{3/2}$ and Fe2p$_{1/2}$, respectively, and the doublet peaks of Fe$^{3+}$-O species at $\sim$711.4 eV and $\sim$725.2 eV for Fe2p$_{3/2}$ and Fe2p$_{1/2}$, respectively. It is to be noted that the composite sample shows no change in the peak positions, except for the Fe$^{3+}$ oxidation state which shows a shift of $\sim$0.6 eV toward higher binding energy. This chemical shift of Fe$^{3+}$ arises due to the variation in the electronegativity of Fe and O in the composite sample. The Fe$^{2+}$:Fe$^{3+}$ ratio turns out to be 45.4:54.6 and 49.4:50.6 in BFO-H and Com-H, respectively.

Figure 5 shows the C 1s core-level spectra for both BFO-H and Com-H. The spectra corresponding to Com-H were fitted by using the contribution of the following six components: (i) defects, (ii) O=C-O, (iii) C=O, (iv) C-C, (v) C=C, and (vi) Fe-C bonds. The peak at 283.9 eV corresponds to the Fe-C bonds. Existence of this peak proves the presence of Fe-C bonds in the Com-H sample. Figure 6 shows the O 1s peak fitted with contribution from the lattice oxygen, oxygen loss, and surface oxygen. Their characteristic peaks appear at lower, intermediate, and higher binding energies. For both the samples, the O 1s peaks were fitted by keeping the peak position, line shape, and the width same. The oxygen loss turns out to be 18.1 and 26.5\%, respectively, for the BFO-H and Com-H.  

We analyzed the C 1s spectra for Com-H quantitatively for determining the concentration of different bonds such as Fe-C, C-C, C=C, O=C-O, O-C-O and compared the results with those obtained from the analysis of the spectra (Fig. 7) for pure GO. The comparison is given in the tabular form in Table II. The concentration of C=C bonds is found to have decreased in Com-H in comparison to that in GO because of the rise in the Fe-C bonds in Com-H. The difference between the concentration of C=C bonds alone in GO and that of C=C and Fe-C bonds ($\sim$3\%) in Com-H is accounted for in the rise in C-C bonds in Com-H. Expectedly, the concentration of O=C-O and O-C-O bonds has decreased in Com-H because of reduction of GO in the composite. Presence of C-C bonds also indicates that not all the surface carbon ions of graphene are bonded to the Fe ions of BiFeO$_3$ particles. Therefore, both bonded and nonbonded BiFeO$_3$ particles coexist in the Com-H sample. As discussed later, relative concentration of bonded BiFeO$_3$ particles plays an important role in governing the magnetoelectric properties of the nanocomposite.

\begin{table}[ht]
{\caption {The relative concentration (in \%) of surface and interface bonds in GO and Com-H.\\}

\begin{tabular}{p{0.6in}p{0.6in}p{0.6in}} \hline\hline
Bonds & GO \newline (\% bonds) & Com-H \newline (\% bonds) \\ \hline
Fe-C & - & 22.69\\
C=C & 35.90 & 9.67\\
C-C & 26.19 & 29.04\\
O-C-O & 26.37 & 23.65\\
O=C-O & 11.54 & 8.20\\
Defects & - & 6.74\\
\hline \hline

\end{tabular}}
\end{table}

\begin{figure}[h!]
 \begin{center}
    \includegraphics[scale=0.30]{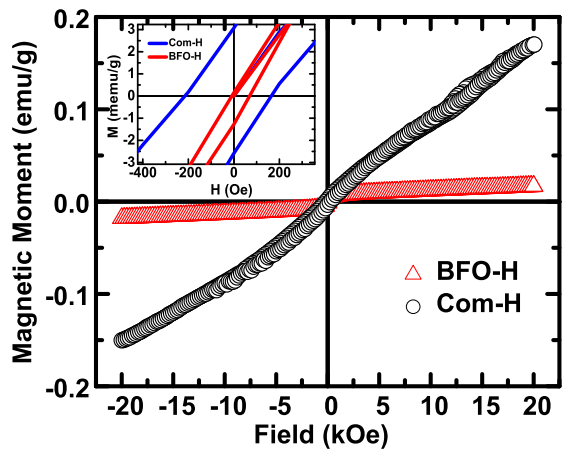} 
    \end{center}
\caption{The room temperature magnetic hysteresis loops for the samples; inset: the portion near the origin has been blown up.}
\end{figure}

\begin{figure}[htp!]
\begin{center}
   \subfigure[]{\includegraphics[scale=0.18]{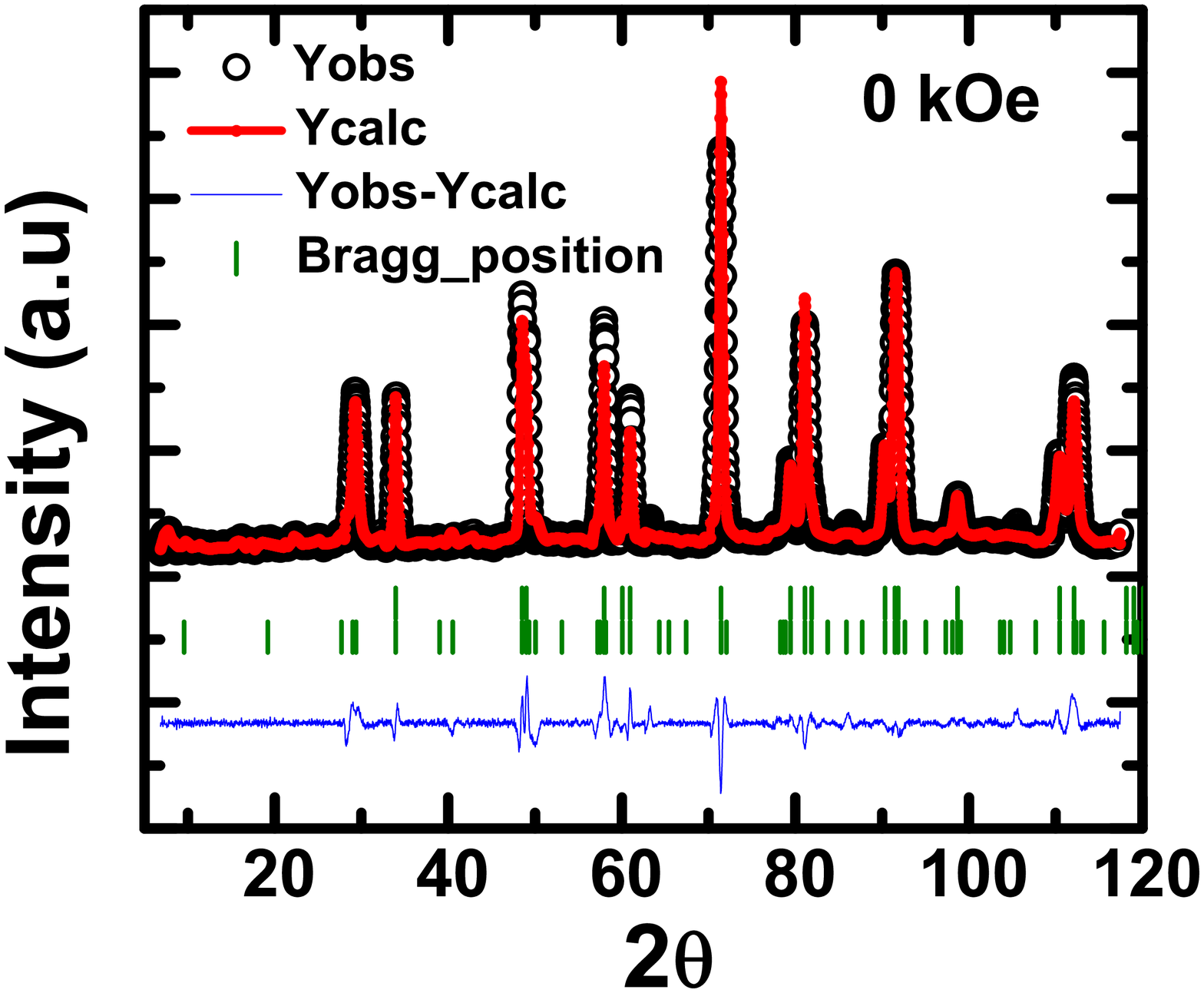}}
   \subfigure[]{\includegraphics[scale=0.18]{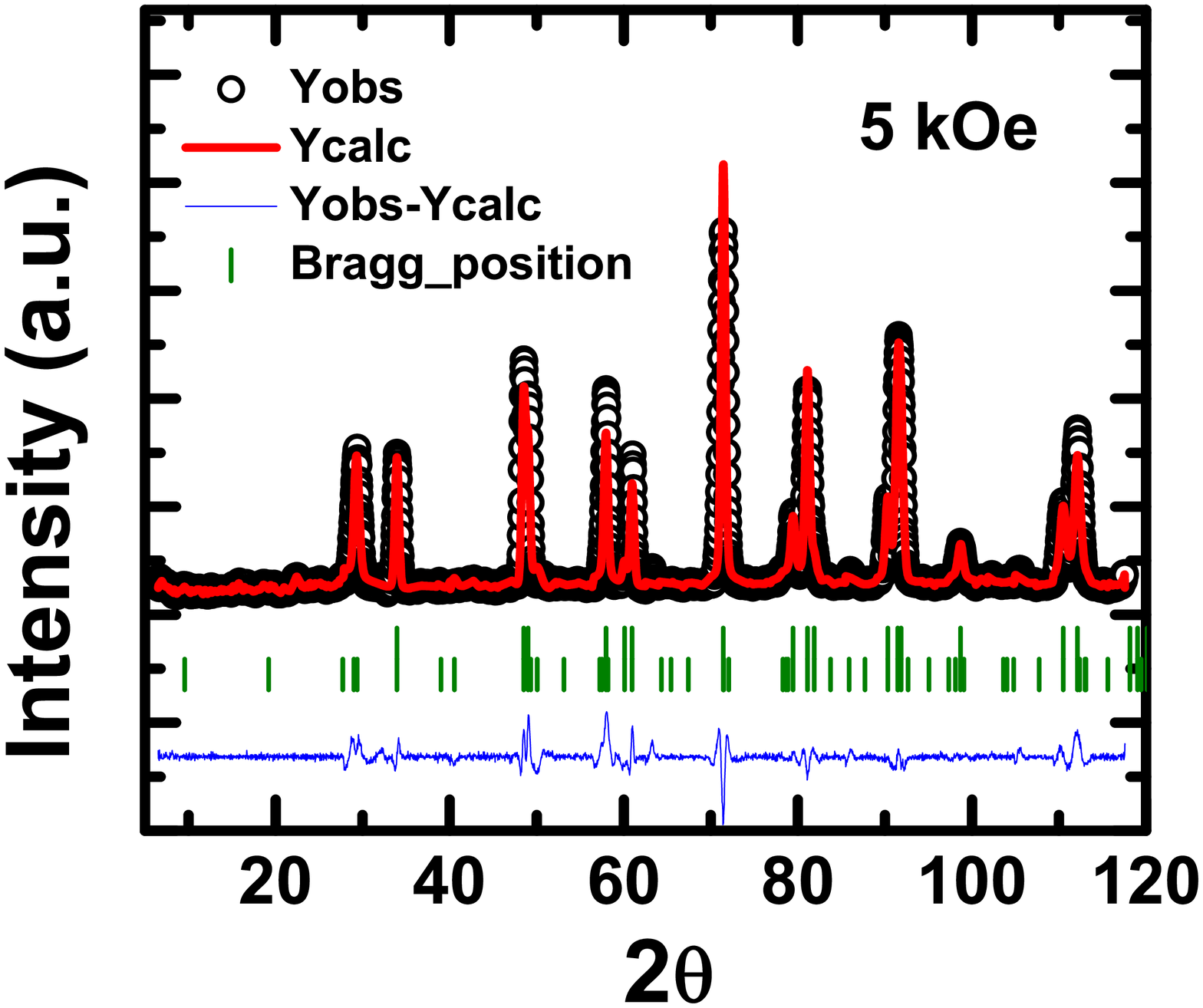}}
   \subfigure[]{\includegraphics[scale=0.18]{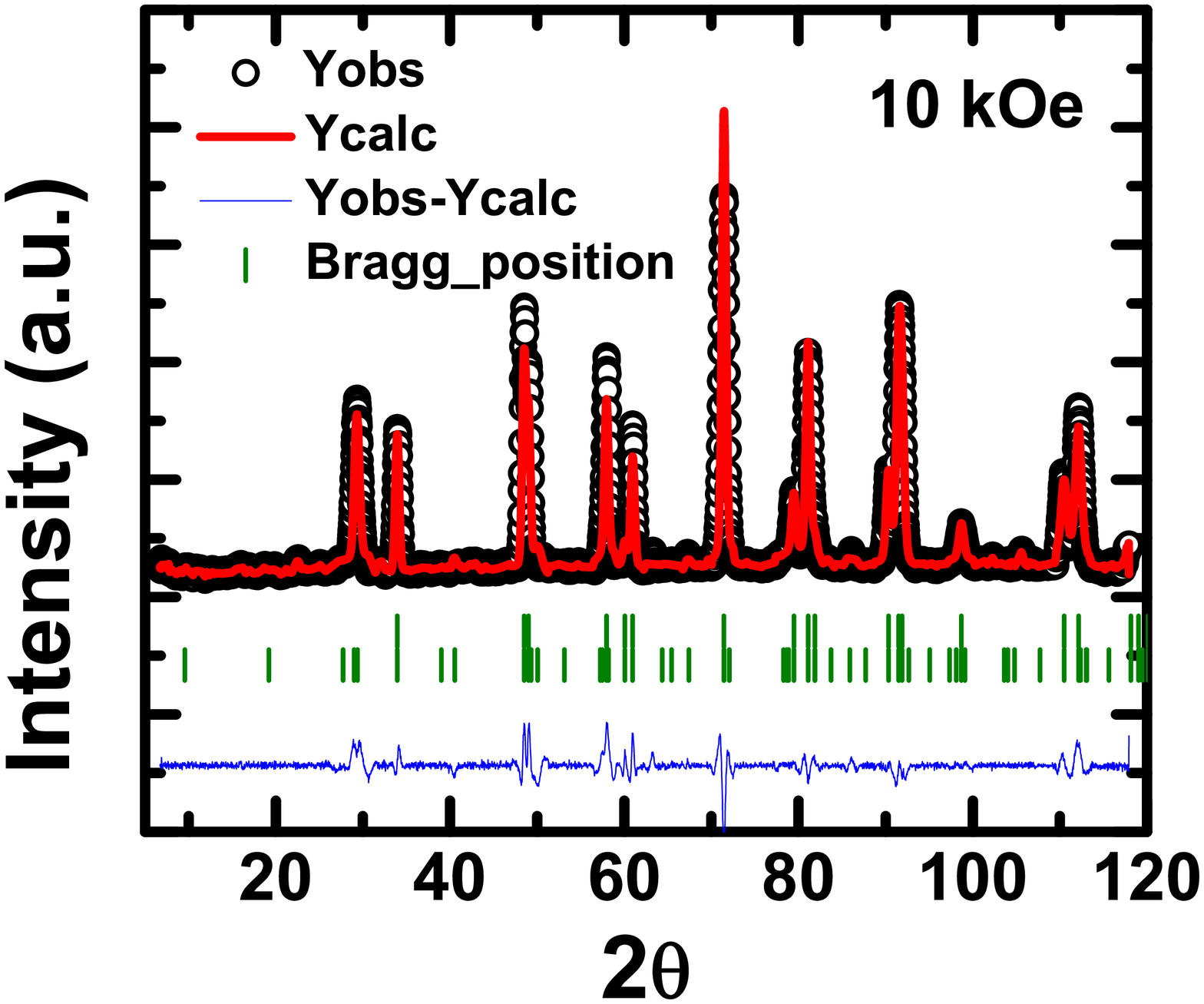}}
   \subfigure[]{\includegraphics[scale=0.18]{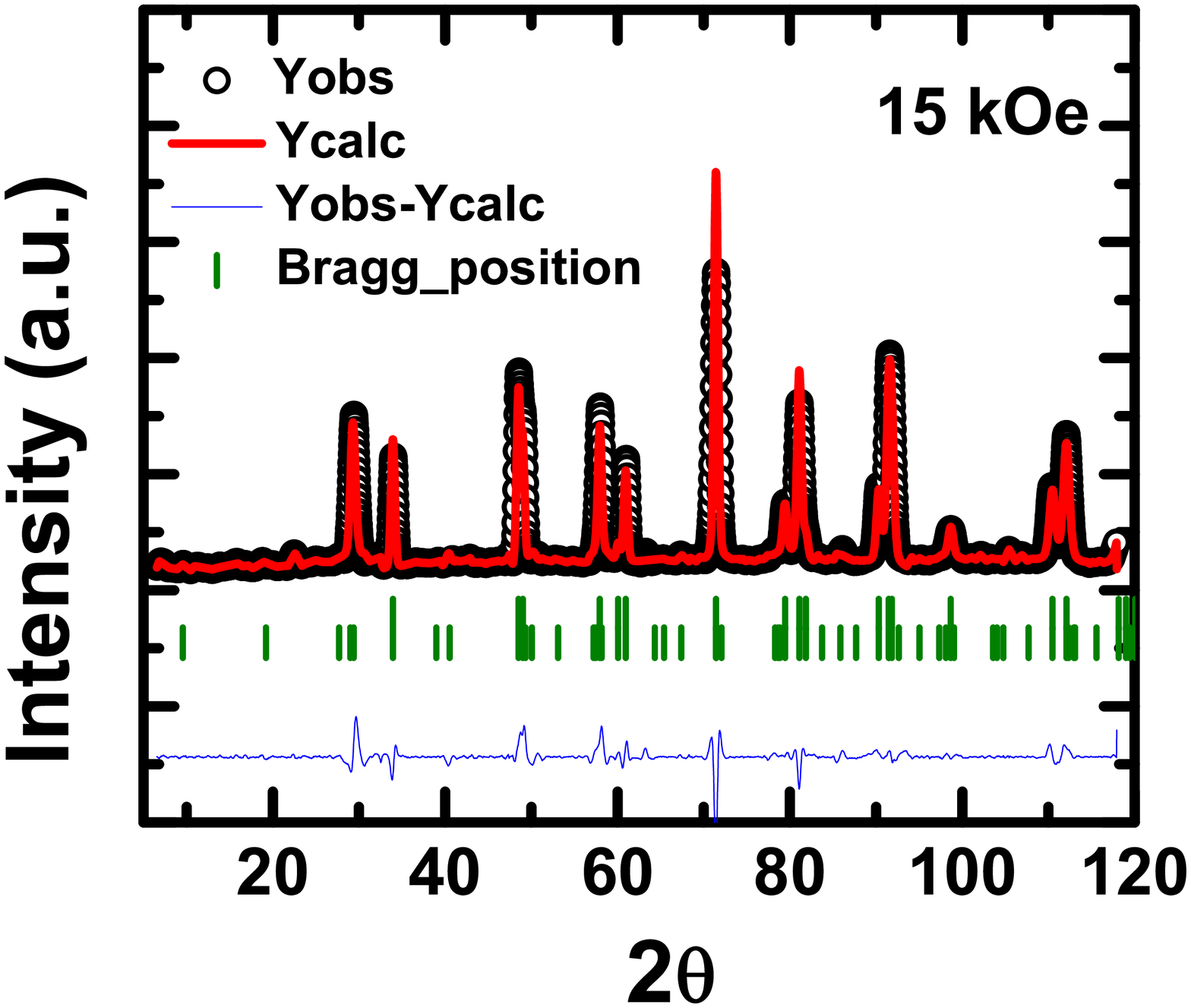}}
   \subfigure[]{\includegraphics[scale=0.18]{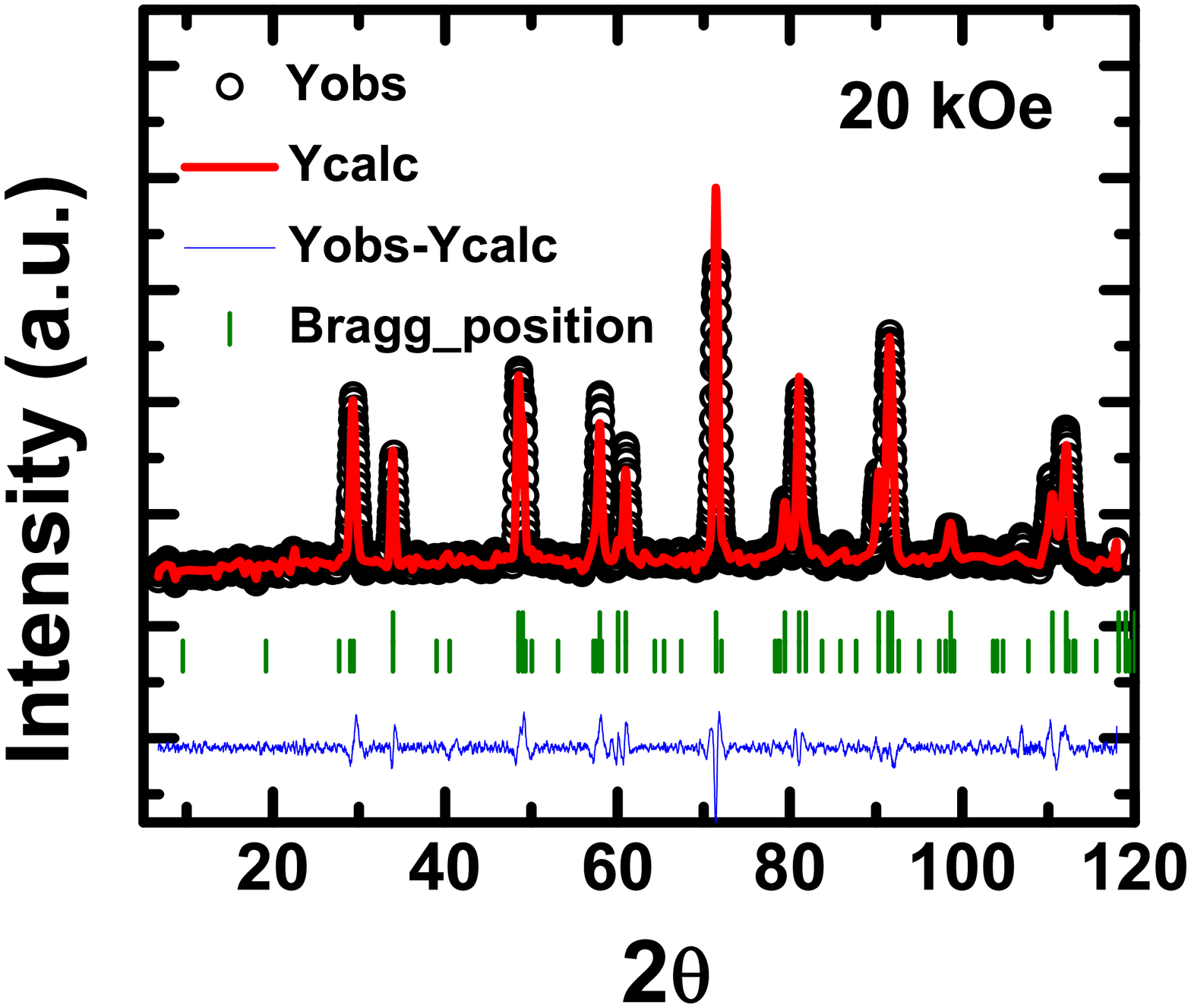}}
   \end{center}
\caption{Powder neutron diffraction data and their refinement by FullProf; R3c and $\Gamma_1$ phases have been considered; the data were recorded at room temperature under different magnetic field.}
\end{figure}

\begin{figure}[h!]
\begin{center}
   \subfigure[]{\includegraphics[scale=0.20]{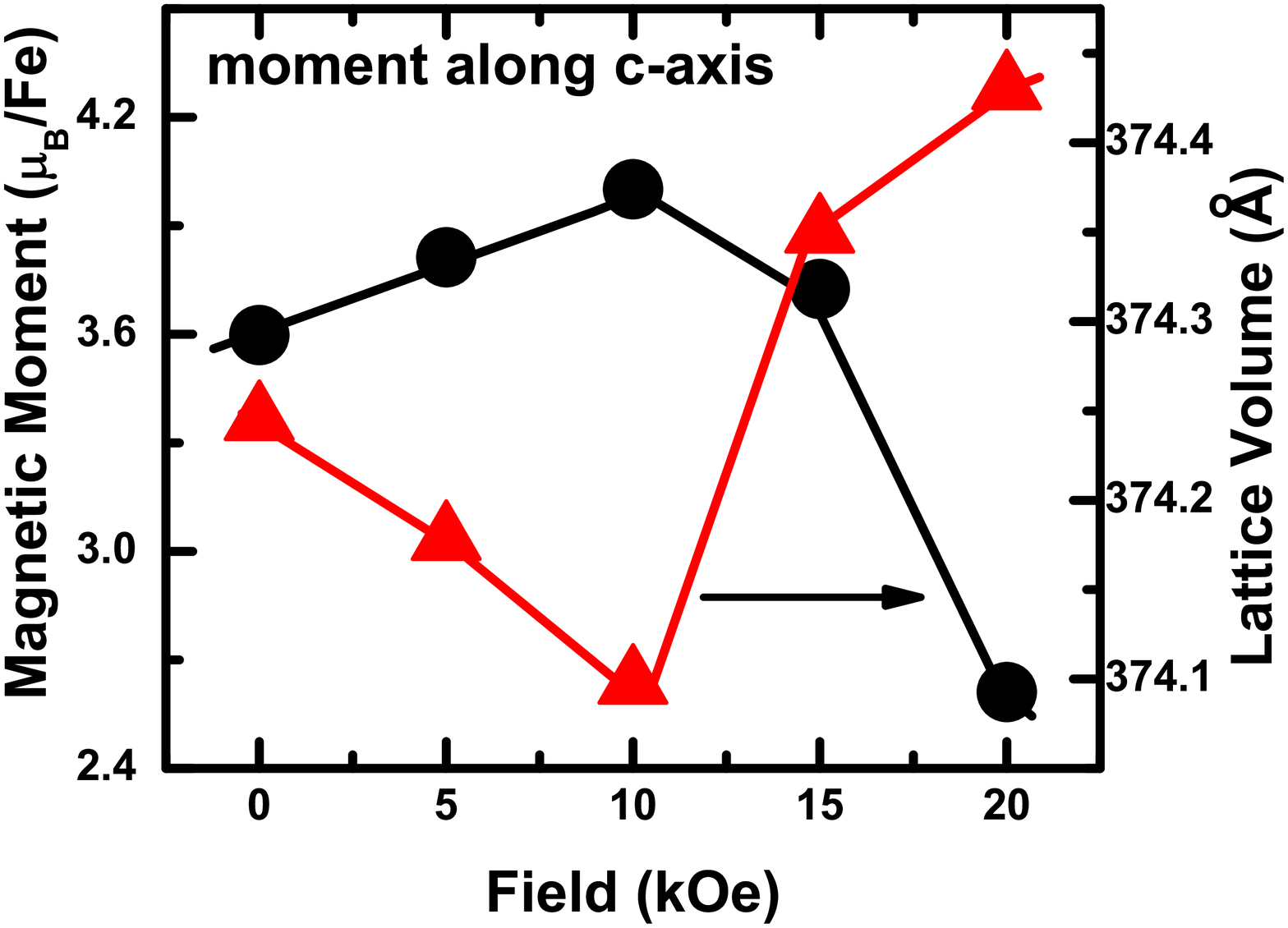}}
   \subfigure[]{\includegraphics[scale=0.20]{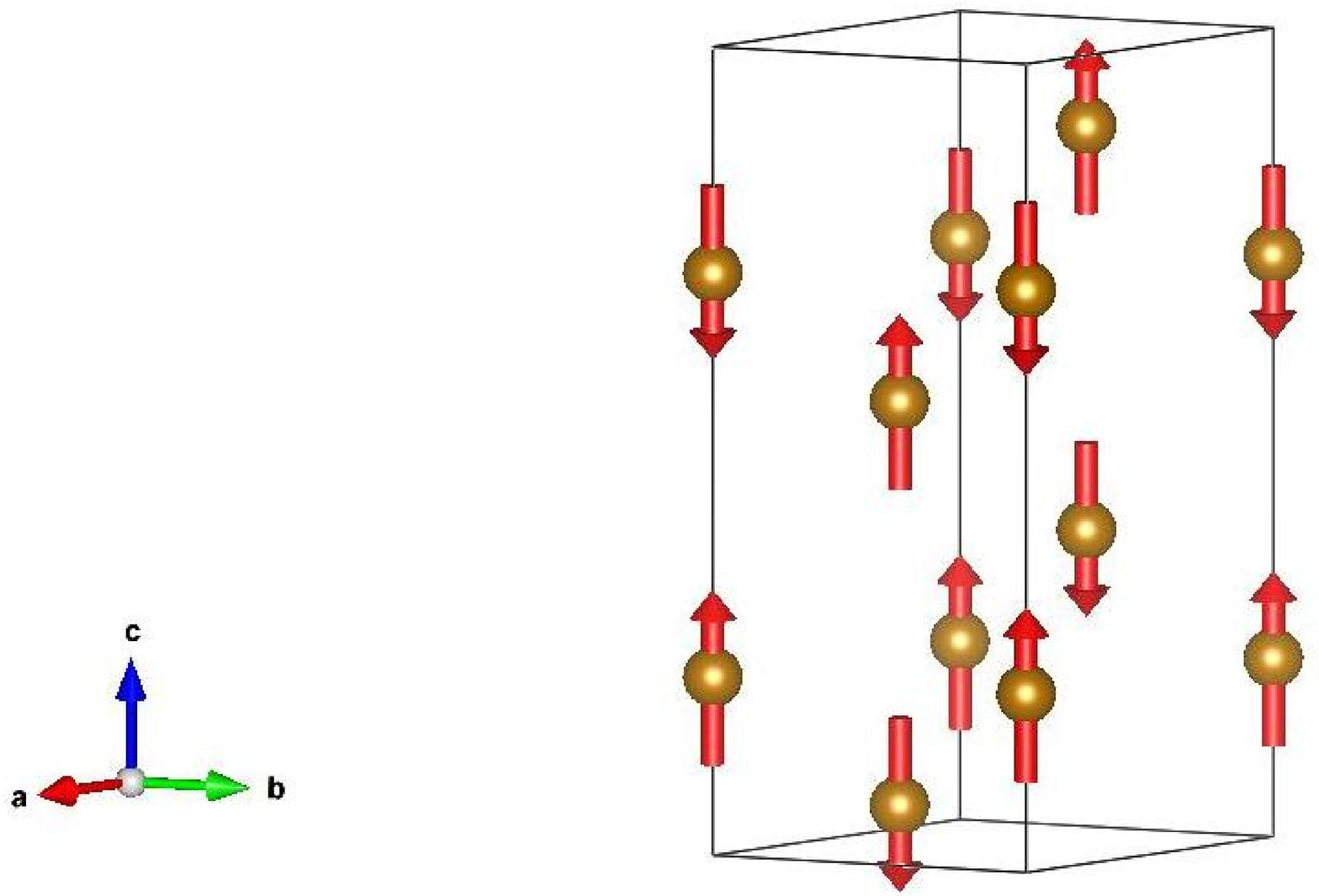}}
   \end{center}
\caption{(a) The variation of the lattice  volume and magnetic moment along c-axis with field obtained from the refinement of neutron diffraction data; (b) the spin structure corresponding to $\Gamma_1$ irreducible representation.}
\end{figure}

\begin{figure*}[htp!]
\begin{center}
   \subfigure[]{\includegraphics[scale=0.25]{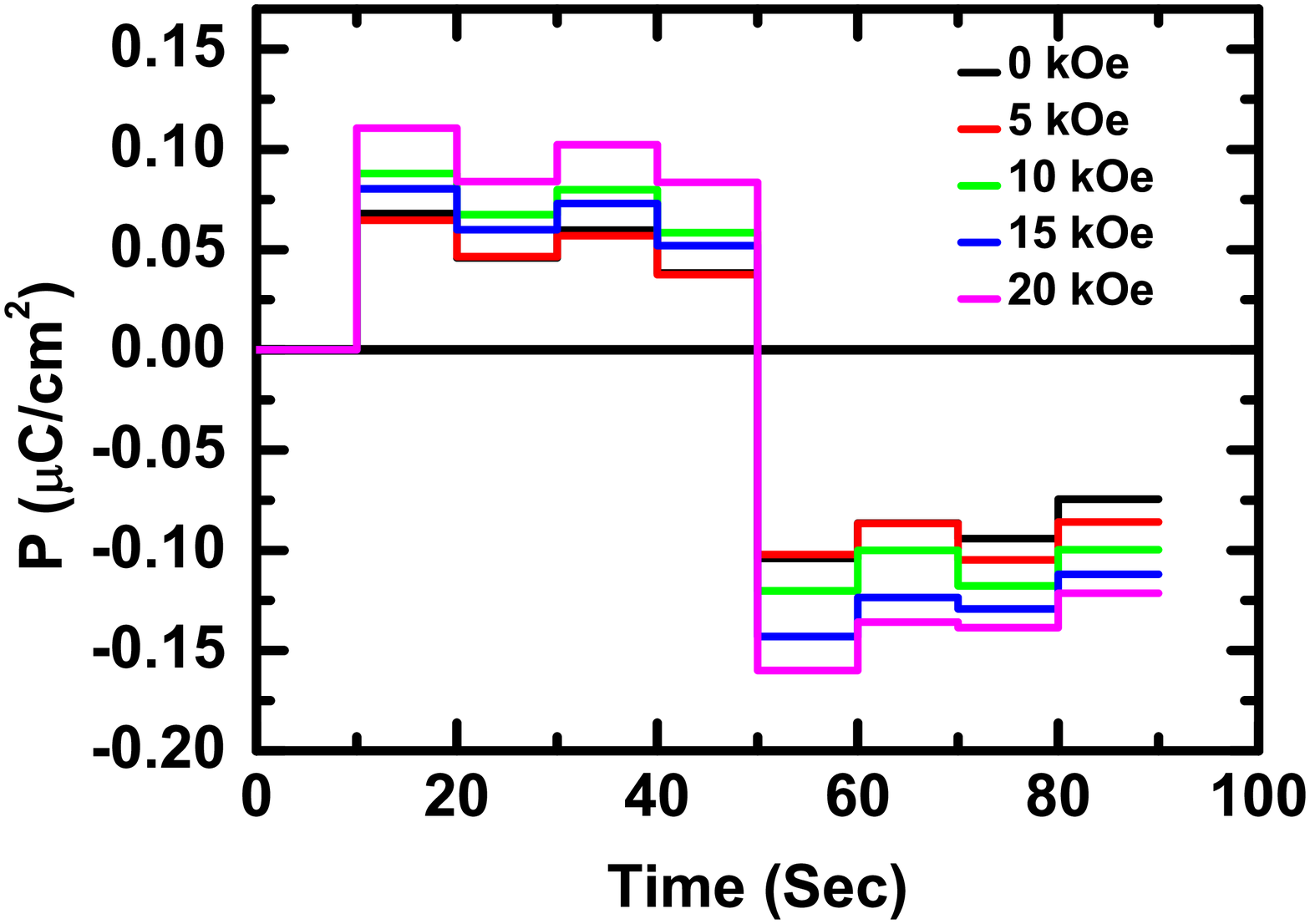}}
   \subfigure[]{\includegraphics[scale=0.25]{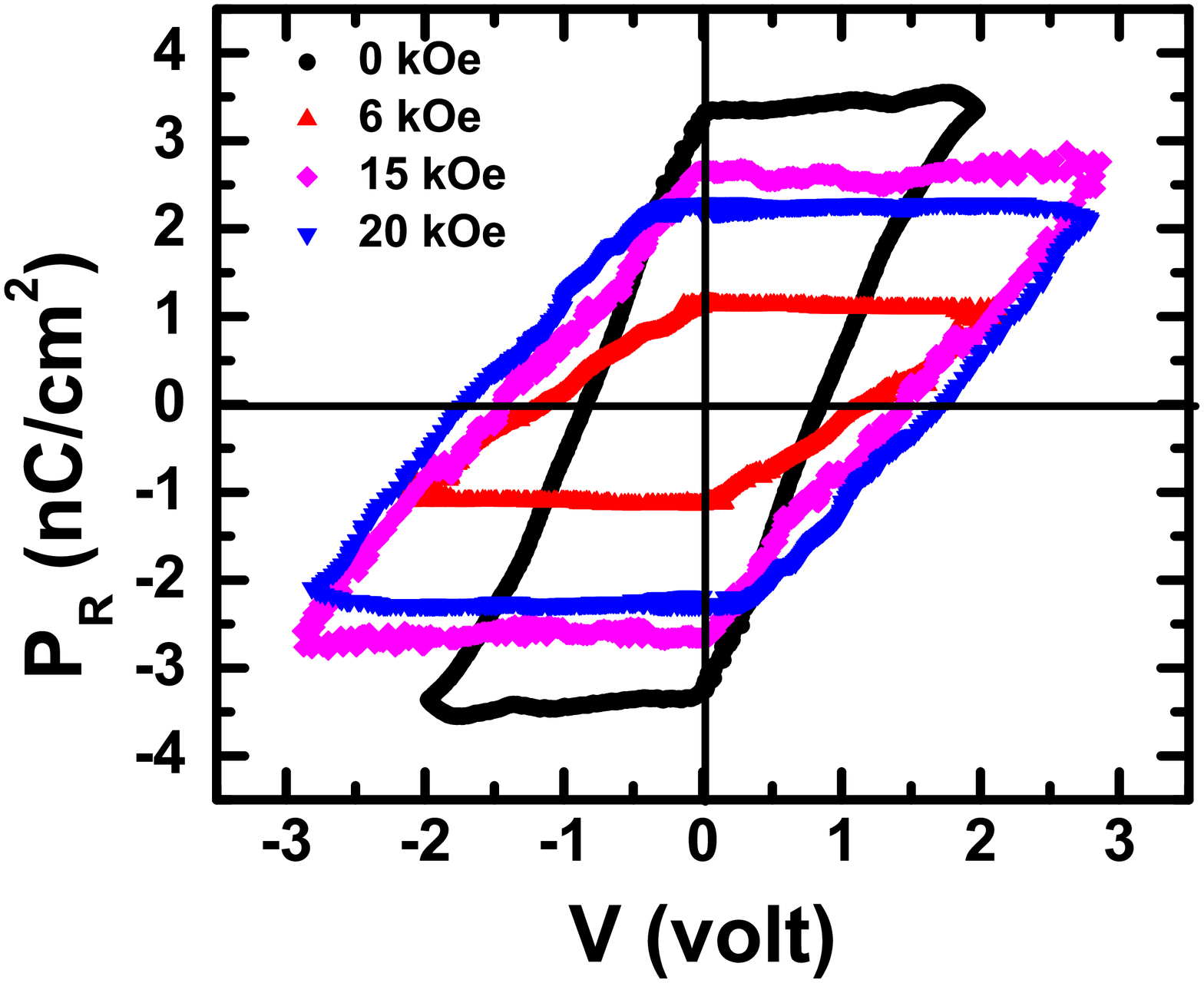}}
   \subfigure[]{\includegraphics[scale=0.25]{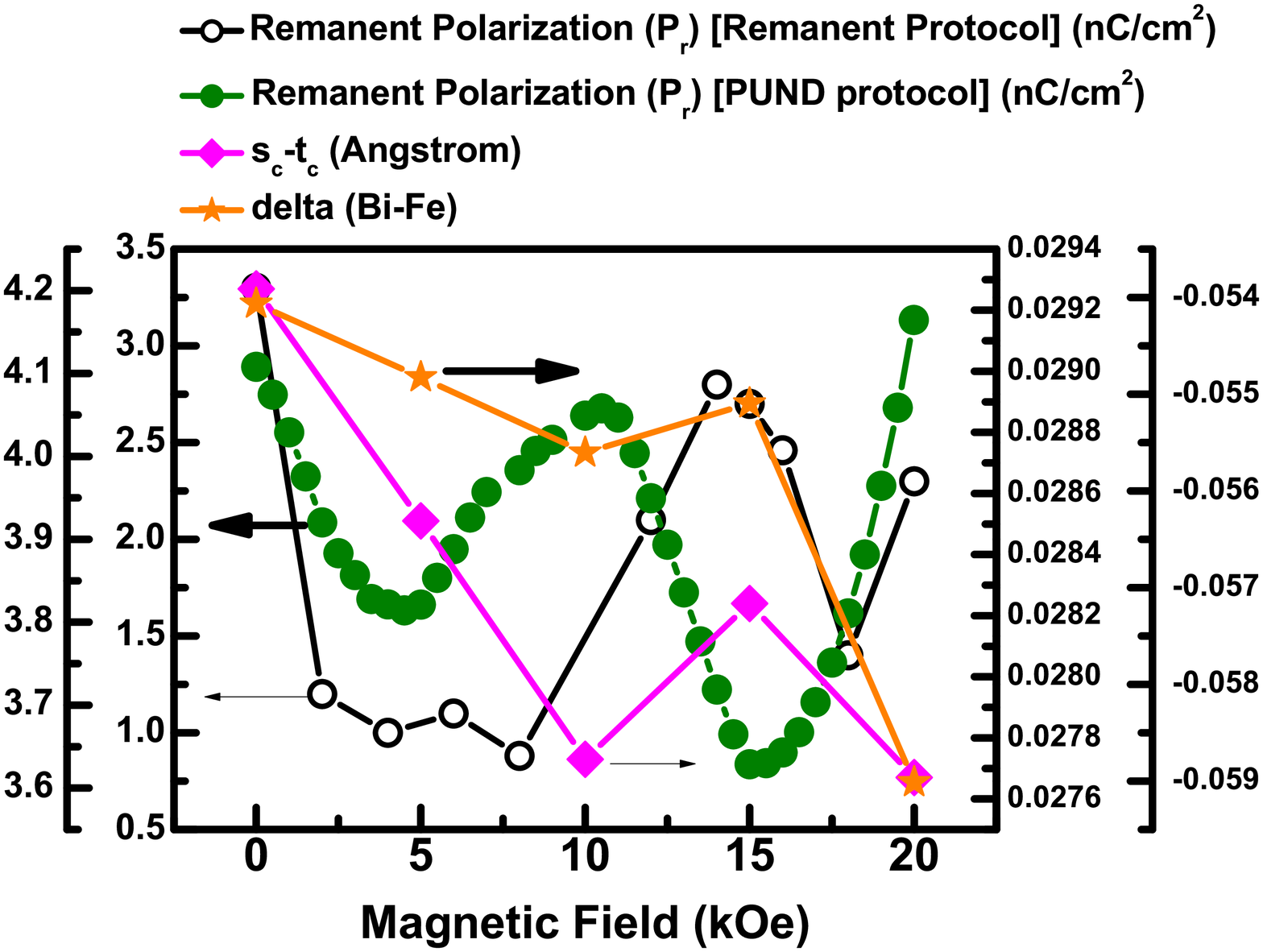}}
   \subfigure[]{\includegraphics[scale=0.25]{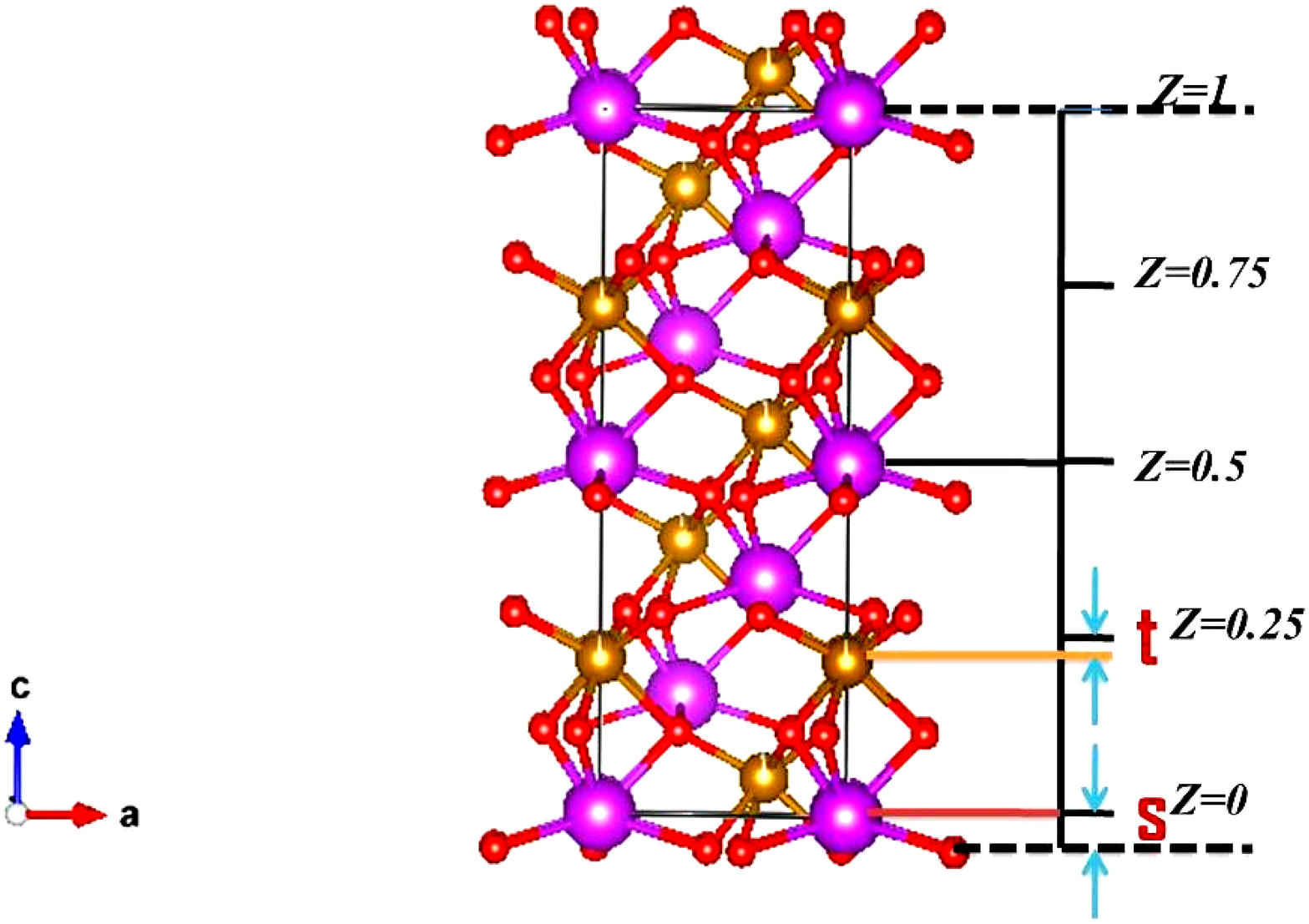}}
   \end{center}
\caption{The remanent ferroelectric polarization under different magnetic fields obtained from the (a) PUND protocol, (b) remanent hysteresis loops; (c) variation of $P_R$ with $H$ obtained from remanent hysteresis loops (inner left axis) and PUND (outer left axis); unit cell off-centered displacement - s$_c$-t$_c$ (inner right axis) and net off-centering with respect to the oxygen cages (outer right axis) - obtained from refinement of powder neutron diffraction data; (d) section of the R3c crystallographic structure is shown in hexagonal setting to highlight the off-centered displacements of Bi and Fe ions along [001].}
\end{figure*}

Summarizing the results obtained from the X-ray diffraction, transmission electron microscopy, IR and Raman spctrometry, and X-ray photoelectron spectroscopy, we notice that (i) the functionalization of RGO by BFO nanoparticles of average size $\sim$20 nm is incomplete as both bonded and nonbonded BFO nanoparticles coexist (defect spacing in RGO/BFO nanocomposite is higher than that in GO which is obtained from greater extent of functionalization of graphene layers by oxygen), (ii) indeed Fe-C bonds have formed at the RGO/BFO interface in the cases where BFO nanoparticles are bonded with RGO in Com-H sample leading to the concomittant rise in covalency in Fe-O bonds, (iii) in spite of lattice distortion (resulting in red and blue shift of characteristic Raman modes), the average global crystallographic structure is still retaining $R3c$ symmtery.

Figure 8 shows the room temperature magnetization ($M$) versus field ($H$) hysteresis loops for Com-H and BFO-H. Contrary to the prediction made \cite{Zanolli-2} in the case of graphene/BaMnO$_3$, the ferromagnetic component is found to have weakened in Com-H. The magnetization ($M$) does not tend to saturate ($M_{20 kOe}$ = 0.017 and 0.17 emu/g, respectively, for BFO-H and Com-H) within the applied field limit ($\pm$20 kOe). The coercivity ($H_C$), of course, turns out to be much larger in Com-H ($H_C$ = 38 and 190 Oe, respectively, for BFO-H and Com-H). It is well known now that, for BFO, the nanoparticles exhibit enhancement in ferromagnetic component because of larger spin canting and incomplete spin spiral (this is effective in particles smaller than $\sim$62 nm - the wavelength of the spin spiral in BiFeO$_3$). It has been shown \cite{Goswami} from powder neutron diffraction experiment that the canting angle could enhance to $\sim$6$^o$ (from $\sim$1$^o$ in bulk sample \cite{Spaldin}) in particles of size $\sim$25 nm. In the present case, variation of the ferromagnetic component could result from change in the charge states of the ions and formation of Fe-C bonds at the RGO/BFO interface in Com-H. The Fe-C exchange coupling interaction (which leads to the exchange splitting of the Dirac bands in graphene layer as well as renormalization of the Fe moments in the interface; exchange field was shown \cite{Song,Wu} to be of the order of 10 to 100T at the BiFeO$_3$/graphene interface) possibly gives rise to the renormalization of the exchange coupling parameters at the bulk as well which, in turn, could give rise to drop in the spin canting angle. The ferromagnetic component, therefore, weakens. Interestingly, in spite of weakening of the ferromagnetic component, the coercivity ($H_C$) turns out to be larger in Com-H. The pinning of domains by defects appears to be stronger in Com-H. Because of the surface ferromagnetism and core antiferromagnetism, small amount of exchange bias field ($H_E$) could also be observed ($H_E$ = 31 and -25 Oe, respectively, for BFO-H and Com-H).

The powder neutron diffraction data were recorded on Com-H sample at room temperature under different magnetic field across 0-20 kOe. We ignored the spin spiral and considered collinear structure. Long spiral length ($\sim$62 nm) in BiFeO$_3$ allows such consideration. The data were refined by FullProf by considering the $R3c$ space group and $\Gamma_1$ irreducible representation (propagation vector k = 0). Group theory analysis of the spin structure for $R3c$ and propagation vector k = 0 yields three irreducible representations $\Gamma_1$, $\Gamma_2$, and $\Gamma_3$ (reducible representation $\Gamma$(Fe, 6a) = $\Gamma_1$ + $\Gamma_2$ + 2$\Gamma_3$). Refinement shows that the $\Gamma_1$ offers the best fitting \cite{supplementary}. The refinement is shown in Fig. 9. Because of the limitations of the current neutron diffraction experiments, we could not detect the change in spin structure in the RGO/BFO nanocomposite with respect to the spin structure of nanoscale BiFeO$_3$. The lattice parameters, magnetic moment, ion positions, and the fit statistics are included in the supplementary document \cite{supplementary}. The variation of the magnetic moment along c-axis ($c_3$) with the applied magnetic field together with the collinear spin structure corresponding to $\Gamma_1$ are shown in Fig. 10. The nonmonotonic field dependence of $c_3$ possibly reflects field-dependent competition between interface and bulk magnetization and consequent renormalization of the average bulk magnetic moment calculated from the powder neutron diffraction by considering collinear G-type spin structure $\Gamma_1$. Interestingly, the lattice volume is found to decrease and then increase (Fig. 10) indicating switch from negative to positive magnetostriction as the applied magnetic field is enhanced across 0-20 kOe. This could be (as discussed later) due to field-dependent competition between bulk and surface/interface magnetic anisotropy.          

We finally discuss the results of the measurement of ferroelectric polarization under different magnetic fields. Both the positive-up-negative-down (PUND) \cite{Feng} and remanent ferroelectric hysteresis loop protocols \cite{Chowdhury} were employed for the measurements. Figure 11 shows the results for Com-H. Under zero magnetic field, the polarization ($P_R$) turns out to be $\sim$4.0 nC/cm$^2$ \cite{supplementary}. Along with the results obtained from direct electrical measurements on the nanoparticle assembly, the net off-centered displacement ($\delta$) in the unit cell, estimated from the powder neutron diffraction data, are also plotted in Fig. 11. Interestingly, both the $P_R$ and $\delta$ exhibit nonmonotonic pattern of variation with the magnetic field ($H$) across 0-20 kOe. $\textit{This is the central result of this paper}$. There is, however, quantitative difference between the $P_R-H$ patterns obtained from direct electrical measurements by PUND and remanent hysteresis protocols. As pointed out in a detailed work \cite{Chowdhury}, PUND protocol contains contribution from nonhysteretic polarization as well. As a result, the intrinsic hysteretic polarization component cannot be accurately extracted by PUND protocol. The close conformation of the $P_R-H$ and $\delta-H$ patterns obtained from remanent hysteresis protocol and neutron diffraction data provides further proof of better efficacy of remanent hysteresis protocol in extracting the intrinsic hysteretic polarization in a sample where contribution from several spurious effects obscures the intrinsic effect. Nonmonotonic nature of $P_R-H$ in RGO/BFO nanocomposite, therefore, is intrinsic. Pristine BiFeO$_3$, on the contrary, exhibits monotonic suppression of $P_R$ with the increase in magnetic field because of negative magnetoelectric coupling \cite{Lee}. In BiFeO$_3$, stabilization of $R3c$ structure yields coupling of polarization and magnetization via coupling between polar distortion and antiferrodistortion of FeO$_6$ octahedra. While polar distortion takes place along [111] (along [001] in hexagonal setting), antiferrodistortive rotation of FeO$_6$ octahedra occurs around [111]. This, in turn, is coupled to the magnetization contained in the (111) plane. Symmetry bars the 180$^o$ switching of the sense of antiferrodistortive rotation of FeO$_6$ octahedra (and hence the magnetization) as a result of 180$^o$ switching of polar distortion in a single step \cite{Spaldin}. However, it has been demonstrated \cite{Ramesh} that indeed the deterministic 180$^o$ switching of magnetic domains takes place (as a result of that of the ferroelectric domains) under electric field in two steps - first, in-plane rotation by 71$^o$ and then out-of-plane rotation by 109$^o$. Piezo- and magnetostriction associated with this complex pathway of domain rotation influences the magnitude of the polarization and magnetization of the domains.

As the magnetic field is swept from zero to a maximum, negative magnetoelectric coupling (arises from influence of negative magnetostriction on piezostriction) in pure BiFeO$_3$ yields decrease in polarization with increase in magnetization \cite{Lee}. Suppression of polarization under magnetic field was observed not just in bulk form of the sample but in nanoscale samples as well \cite{Goswami-1}. The bulk and surface magnetocrystalline anisotropy ($K_B$ and $K_S$) results in negative bulk and surface magnetostriction ($\lambda_B$ and $\lambda_S$) in nanoscale BiFeO$_3$ assuming $R3c$ structure. Negative magnetostriction leads to negative magnetoelectric coupling, i.e., suppression of polarization under magnetic field. In an assembly of single crystalline nanoparticles, averaging over the domain configurations and the properties, therefore, yields similar decrease (increase) in polarization (magnetization) at any given state. This magnetostrictive suppression of piezostriction, observed in pure BiFeO$_3$, can be altered and consequent enhancement of magnetostriction driven polarization is possible. In BiFeO$_3$/reduced-graphene-oxide nanocomposite, emergence of Fe-C bonds (from strong hybridization of $3d_{x^2-y^2}/3d_{z^2}$ orbitals of Fe and $p_z$ orbitals of C) and consequent magnetization at the graphene-BiFeO$_3$ interface regions opens up this possibility. It has already been shown \cite{Wu,Niu} that the Fe-C bonding leads to the generation of large exchange field (of the order of 10 to 100T) which gives rise to the split of Dirac bands in the graphene layer as well as proximity induced magnetization. This exchange field alters the magnetization at the interface reflected in the enhancement of the antiferromagnetic component in RGO/BFO nanocomposite. Change in the magnetic moment at the interface was earlier predicted by theoretical calculations \cite{Ho}. The anisotropy ($K_I$) associated with this interface magnetization could yield positive surface magnetostriction. This, in turn, could give rise to the rise in $P_R$ with $H$, i.e., positive magnetoelectric coupling. It has been shown \cite{Labaye} earlier that indeed the ratio of surface/interface anisotropy and bulk anisotropy ($K_I$/$K_B$) governs the spin structure significantly. Different structure emerges as $K_I$ or $K_S$ enhances with respect to $K_B$. Conversely, different spin structures at the interface could yield different $K_I$ and, as a consequence, positive interface magnetostriction. Field-dependent switch from negative to positive magnetostriction has indeed been observed in the present case. The mapping of lattice volume with the applied magnetic field (Fig. 10) shows decrease and then increase in lattice volume as the field is swept across 0-20 kOe. This observation underlines the important role of interface magnetization driven by Fe-C bonds. Of course, how $K_I$, in the present case, yields positive interface magnetostriction and the ratio $K_I$/$K_B$ governs the influence of magnetostriction on piezostriction as a function of applied magnetic field ($H$) - from negative to positive - needs to be studied in detail separately.

\begin{figure}[ht!]
 \begin{center}
    \includegraphics[scale=0.20]{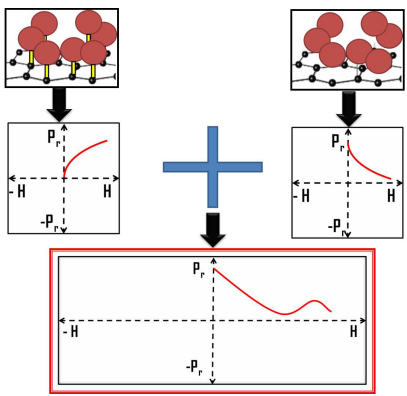} 
    \end{center}
\caption{The schematic of the bonded and nonbonded BiFeO$_3$ particles with reduced graphene oxide layers; the bonded particles because of the presence of Fe-C bonds and exchange coupling interactions across them could offer positive magnetostrictive magnetoelectric coupling; the nonbonded ones, on the other hand, would exhibit negative magnetoelectric coupling; field-dependent competition between these two fractions is eventually yielding the nonmonotonic variation of $P_R$ with $H$. Tuning of the volume fractions of the bonded and nonbonded BiFeO$_3$ particles, therefore, offer a pathway to tune the swithcing of magnetoelectric coupling - from purely negative to mixed positive and negative to purely positive.}
\end{figure}

Based on the above discussion, in Fig. 12, we present the schematic of the model which describes the physics behind the observed nonmonotonic $P_R-H$ pattern. The Com-H sample contains both bonded (via Fe-C bonds) and nonbonded (attached via van der Waals bonds) BiFeO$_3$ particles. While magnetostriction due to symmetric exchange interaction (i.e. the exchange interaction which does not involve spin-orbit coupling) across Fe-O-Fe bonds, in the absence (relevant for BFO-H) and presence of proximity coupling via formation of Fe-C bonds (relevant for Com-H because of the presence of these bonds at the interface regions), yields, respectively, negative and positive magnetoelectric coupling, the field-dependent competition (shown in Fig. 12) between the positive and negative magnetoelectric coupling in bonded and nonbonded BiFeO$_3$ particles yields the observed nonmonotonic pattern of variation of $P_R$ with $H$. The $P_R$ decreases initially and then rises and finally decreases again as the applied magnetic field is swept from zero to $\sim$20 kOe. Relative volume fraction of the bonded and nonbonded BiFeO$_3$ nanoparticles governs the relative strength of increase and decrease of the $P_R$ as a function of $H$. Coexistence of positive and negative magnetostrictive regions and engineering of their respective volume fractions have earlier been demonstrated \cite{Gou} within the bulk of the Fe-Ga alloys. This was projected to yield field-dependent switch in ferroelectric polarization in ferroelectric/Fe-Ga alloy multilayer composites. In the present case, reconstruction of interface in reduced-graphene-oxide/BiFeO$_3$ nanocomposite is shown to be actually yielding the result projected in Ref. 27. Engineering of the bulk and interface anisotropies and hence of the bulk and interface magnetostrictions further could yield eventually an interesting custom-designed periodic nonmonotonic pattern of variation of $P_R$ with $H$. Such custom-designed $P_R-H$ pattern will obviously expand the horizon of the nanospintronic applications by a great extent. 

\section{Conclusion}

In conclusion, the RGO/BFO nanocomposite exhibits a range of interesting properties such as decrease in ferromagnetic component yet rise in coercivity and finite negative exchange bias and nonmonotonic dependence of ferroelectric polarization on applied magnetic field. This ensemble of unusual properties results from reconstruction of BiFeO$_3$/graphene interface via formation of Fe-C bonds and could find interesting applications. It could also trigger fresh research on atomic scale engineering of the interfaces in the composites or heterostructures for far enhanced functionalities. 

\begin{center}
\bf{ACKNOWLEDGMENTS}
\end{center}

One of the authors (T.C.) acknowledges DST-INSPIRE research fellowship. Another author (A.M.) acknowledges support from Science and Engineering Research Board (SERB), Government of India (Grant No. CRG/2021/002132).

\end{document}